\documentstyle[12pt]{article}
\catcode`\@=11

\catcode`\@=11

\def\section{\@startsection{section}{1}{\z@}{3.5ex plus 1ex minus
   .2ex}{2.3ex plus .2ex}{\large\bf}}

\def\eqnarray{\let\@currentlabel=\theequation\refstepcounter{equation}
    \global\@eqnswtrue
    \global\@eqcnt\z@\tabskip\@centering\let\\=\@eqncr
    $$\halign to \displaywidth\bgroup\@eqnsel\hskip\@centering
      $\displaystyle\tabskip\z@{##}$&\global\@eqcnt\@ne
       \hfil${{}##{}}$\hfil
      &\global\@eqcnt\tw@ $\displaystyle\tabskip\z@{##}$\hfil
       \tabskip\@centering&\llap{##}\tabskip\z@\cr}
\def\lefteqn#1{\hbox to 4\arraycolsep{$\displaystyle #1$\hss}}
\def\thesection{\arabic{section}}

\def\appendix{\setcounter{section}{0}
        \def\thesection{Appendix.}
        \def\theequation{\Alph{section}.\arabic{equation}}}
\long\def\@makefntext#1{\parindent 0cm\noindent
\hbox to 1em{\hss$^{\@thefnmark}$}#1}
\def\IR{{\hbox{{\rm I}\kern-.2em\hbox{\rm R}}}}
\def\IH{{\hbox{{\rm I}\kern-.2em\hbox{\rm H}}}}
\def\IC{{\ \hbox{{\rm I}\kern-.6em\hbox{\bf C}}}}
\def\IZ{{\hbox{{\rm Z}\kern-.4em\hbox{\rm Z}}}}

\newcommand{\beq}{\begin{equation}}
\newcommand{\eeq}{\end{equation}}
\begin{document}
%
%
%
%
\def\citen#1{%
\edef\@tempa{\@ignspaftercomma,#1, \@end, }
\edef\@tempa{\expandafter\@ignendcommas\@tempa\@end}%
\if@filesw \immediate \write \@auxout {\string \citation {\@tempa}}\fi
\@tempcntb\m@ne \let\@h@ld\relax \let\@citea\@empty
\@for \@citeb:=\@tempa\do {\@cmpresscites}%
\@h@ld}
%
\def\@ignspaftercomma#1, {\ifx\@end#1\@empty\else
   #1,\expandafter\@ignspaftercomma\fi}
\def\@ignendcommas,#1,\@end{#1}
%
%
\def\@cmpresscites{%
 \expandafter\let \expandafter\@B@citeB \csname b@\@citeb \endcsname
 \ifx\@B@citeB\relax 
    \@h@ld\@citea\@tempcntb\m@ne{\bf ?}%
    \@warning {Citation `\@citeb ' on page \thepage \space undefined}%
 \else
    \@tempcnta\@tempcntb \advance\@tempcnta\@ne
    \setbox\z@\hbox\bgroup 
    \ifnum\z@<0\@B@citeB \relax
       \egroup \@tempcntb\@B@citeB \relax
       \else \egroup \@tempcntb\m@ne \fi
    \ifnum\@tempcnta=\@tempcntb 
       \ifx\@h@ld\relax 
          \edef \@h@ld{\@citea\@B@citeB}%
       \else 
          \edef\@h@ld{\hbox{--}\penalty\@highpenalty \@B@citeB}%
       \fi
    \else   
       \@h@ld \@citea \@B@citeB \let\@h@ld\relax
 \fi\fi%
 \let\@citea\@citepunct
}
%
\def\@citepunct{,\penalty\@highpenalty\hskip.13em plus.1em minus.1em}%
%
%
\def\@citex[#1]#2{\@cite{\citen{#2}}{#1}}%
%
%
\def\@cite#1#2{\leavevmode\unskip
  \ifnum\lastpenalty=\z@ \penalty\@highpenalty \fi 
  \ [{\multiply\@highpenalty 3 #1
      \if@tempswa,\penalty\@highpenalty\ #2\fi 
    }]\spacefactor\@m}
\let\nocitecount\relax  
%

\begin{titlepage}
\vspace{.5in}
\begin{flushright}
\vspace{.2in}
WATPHYS TH-98/02\\
gr-qc/9806004\\
March 1998\\
\end{flushright}
\vspace{1in}      
\begin{center}
{\Large\bf
A New Approach to Black Hole Microstates\footnote{This essay 
received an ``honorable mention'' from the
Gravity Research Foundation, 1998.}}\\
\vspace{.4in}
R{\sc ichard}~J.~E{\sc pp}\footnote{\it email: epp@avatar.uwaterloo.ca}
{\sc and} R.~B.~M{\sc ann}\footnote{\it email: mann@avatar.uwaterloo.ca}\\
       {\small\it Department of Physics}\\
       {\small\it University of Waterloo}\\
       {\small\it Waterloo, Ontario N2L 3G1}\\{\small\it Canada}\\
\end{center}

\vspace{.5in}
\begin{center}
\begin{minipage}{5in}
\begin{center}
{\large\bf Abstract}
\end{center}
{\small
If one encodes the gravitational degrees of freedom in an
orthonormal frame field there is a very natural first order action
one can write down
(which in four dimensions is known as the Goldberg action).  
In this essay we will show that this action contains a 
boundary action for certain microscopic degrees of freedom living
at the horizon of a black hole, 
and argue that these degrees of freedom hold great promise for
explaining the microstates responsible
for black hole entropy, in any number of spacetime dimensions.
This approach faces many interesting challenges, both
technical and conceptual.
}
\end{minipage}
\end{center}
\end{titlepage}
\addtocounter{footnote}{-3}

In the 1970's we were introduced to the fact that black holes are
thermodynamic objects, whose entropy and other thermodynamic
parameters are identified with simple geometrical aspects of the
black hole \cite{Bard,Beke,Hawk4}.  
An example is the celebrated Bekenstein-Hawking
entropy formula $S=A/4$, where $S$ is the entropy and $A$ the area of 
the bifurcation sphere in Planck units.
However, thermodynamics is only an approximation
to a more fundamental description based on statistical mechanics.
In particular, entropy is associated with the number, $\Gamma$, 
of ``microstates" compatible with a given ``macrostate" through 
Boltzmann's formula $S=\ln\Gamma$.  
A puzzle then arises in that general relativity
gives us the macrostate (the black hole, described by just three
parameters: mass, angular momentum, and charge), but apparently
says nothing about the nature of the typically 
huge number $e^S$ of microstates
which presumably must exist to account for the entropy.

Since its recognition this problem has received an enormous amount of 
attention, for two reasons: it is a paradox, expedient for progress in
theoretical physics, and an important one: the 
Bekenstein-Hawking entropy formula contains Planck's constant
and so surely the resolution of this problem will provide insight
into a theory of quantum gravity---currently a holy grail of 
theoretical physics.  Over the years much has been learned, 
and recently string theory
has provided some exciting answers \cite{Horo}, but it is not our purpose
here to review the literature.  Instead we will introduce some ideas,
very geometrical and topological in flavor,
which we believe hold great promise towards providing a 
satisfactory explanation of black hole microstates in any number of
spacetime dimensions
(within the context of general relativity
proper, but containing numerous hints at hidden connections 
with string theory).

These ideas are inspired by two central themes which stand out in the
literature.
The first is that black hole entropy is intimately connected with
topological considerations, as first emphasized 
by Gibbons and Hawking in 1979 \cite{Gibb2}.  
Perhaps the strongest statement of this fact are recent arguments
which suggest that the correct entropy formula, generalized
to encompass arbitrary topology, is \cite{Libe}
\begin{equation}
S={\chi\over 8}A\, ,
\label{generalized entropy}
\end{equation}
where $\chi$ is the Euler number of the black hole.
The second is the very plausible idea that the 
sought for microscopic degrees of freedom 
reside at the horizon, interpreted as a ``boundary'' 
of sorts; after all, the
entropy is proportional to the horizon area.  This type of idea
was first implemented successfully by Carlip in 1995
for black holes in
three dimensions \cite{Carl}, and has since enjoyed other successes
(recent examples include \cite{Sfet,Asht}).
One might be uncomfortable thinking of the horizon
as a boundary, or might wonder ``What about the microstates
which would presumably reside also at the boundary at infinity?''
These concerns are neatly dealt with in the approach we will now
outline.

To begin, on a $d$-dimensional manifold, $M$, encode the gravitational
degrees of freedom not in the metric, but rather in its ``square
root'': an orthonormal frame field composed of a set
of one-form fields $e^a$, or their dual vector fields, $e_a$.  All of
our considerations will then follow from the simple identity
\begin{equation}
\epsilon\, R=\omega^{a}_{\;\;b}\wedge\omega^{b}_{\;\;c}\wedge
\epsilon^{c}_{\;\;a}-d\, (\omega^{a}_{\;\;b}\wedge\epsilon^{b}_{\;\;a})
\label{epsilon R}
\end{equation}
expressing the Ricci scalar density on the left hand side in terms
of the metric compatible torsion-free spin connection, $\omega^{a}_{\;\;b}$,
and the $(d-2)$-form $\epsilon_{ab}=i_{e_b}i_{e_a}\epsilon$.  (Frame
indices are raised and lowered as usual with the frame signature metric.)
It is immediately obvious that the $\omega\wedge\omega\wedge\epsilon$ term
is the natural choice for a first order Lagrangian density, suitable
to yield the local vacuum Einstein equations.\footnote{This was first
observed by Goldberg in the $d=4$ Lorentzian case, who worked in the
context of Sparling forms and their interesting association with
a certain energy-momentum pseudotensor and superpotential \cite{Gold}.}
But when we integrate over $M$ to form the action an interesting
global consideration arises with regard to the boundary, $\partial M$.
To illustrate this most simply, for $d=2$ (\ref{epsilon R}) reduces
to $\epsilon\, R=d\, (2\omega^{01})$.
So for the flat Euclidean disk,
for example, ($ds^2 =dx^2 +dy^2 =dr^2 +r^2 \,d\phi^2$)
we require $\int_{\partial M}\omega^{01}=0$.  For the regular gauge
choice $e^0 =dx$ and $e^1 =dy$, $\omega^{01}=0$ and $\partial M$ is
just the boundary at infinity.  But for the {\it singular} gauge choice
$e^0 =dr$ and $e^1 =r\,d\phi$, $\omega^{01}=d\phi$
and we are forced to augment $\partial M$ with an additional
circle about the origin, i.e. excise this singular point.

More generally, consider a Euclidean black hole, topologically
$\IR ^2 \times\Sigma$ (typically $\Sigma$ is the $(d-2)$-sphere).
A singular gauge choice analogous to the disk example is {\it natural}
in this context: in the ``$\IR^2$ sector''
$e_1$ is chosen to correspond to a Euclidean time flow unit vector field
(with fixed points comprising a bifurcation surface, $\Sigma_0$);
and with $e_0 =n$, the unit normal on $\partial M$, it is easy to show
that
\begin{equation}
\label{gauge-fixed action}
{1\over{2\kappa}}\int_{M}\omega^{a}_{\;\;b}\wedge\omega^{b}_{\;\;c}\wedge
\epsilon^{c}_{\;\;a}
={1\over{2\kappa}}\int_{M}\epsilon\, R+
{1\over\kappa}\int_{\partial M}i_n \epsilon\, K\, ,
\end{equation}
which is the usual gravity action with trace extrinsic curvature
boundary term appropriate for holding fixed,
in the action principle, the boundary $(d-1)$-geometry.
In analogy with the disk example,
here we are forced to excise the bifurcation surface, and
$\partial M=B_{\infty}\cup B_0$, where $B_\infty$ is the boundary at
infinity and $B_0$ is the boundary of a ``thickened bifurcation surface''
(both topologically $S^1 \times \Sigma$).\footnote{In case $\Sigma$ is
not parallelizable (e.g. $\Sigma=S^2$) there are additional points at
which the frame must be multivalued, and that must be excised, introducing
an additional boundary component $B_\star$.  A nonempty $B_\star$ may
have important topological ramifications beyond the scope of this essay,
but in any case it is easily argued that nontrivial microstates do not
reside on $B_\star$, and so henceforth $B_\star$ will be ignored.}

We now introduce the frame rotation degrees of freedom: set 
$e^a =U^{a}_{\;\;b}\hat{e}^b$, where $\hat{e}^b$ is a frame gauge-fixed as
discussed above, and
$U^{a}_{\;\;b}\in SO(d)$.  Employing an obvious matrix notation,
the action in (\ref{gauge-fixed action}) is then augmented on the
right hand side with an additional {\it boundary action}
\begin{equation}
\label{boundary action}
I_{B}[U;\hat{e}]={1\over{2\kappa}}\int_{\partial M}
{\textstyle\rm Tr}\, (U^{-1}dU\wedge \hat{\epsilon})\, ,
\end{equation}
where $\hat{\epsilon}_{ab}=i_{\hat{e}_b}i_{\hat{e}_a}\epsilon$.
With fixed boundary $(d-1)$-geometry (effectively equivalent to
fixed $\hat{\epsilon}\downarrow_{\partial M}$, where 
$\downarrow_{\partial M}$ denotes pullback to $\partial M$), 
the action principle then yields the usual
vacuum Einstein equations in $M$ plus the (highly nontrivial) boundary 
Euler-Lagrange equations 
$d\,(U\hat{\epsilon}U^{-1})\downarrow_{\partial M}=0$.  
Thus, in the spirit of Carlip's 
approach \cite{Carl,Carl4,Carl3},
the would-be gauge degrees of freedom $U$ become physical dynamical
fields on $\partial M$ and (we are suggesting) are the microscopic
degrees of freedom giving rise to the microstates responsible for
black hole entropy; and $\hat{\epsilon}\downarrow_{\partial M}$
encodes information about the black hole macrostate.

With $\partial M=B_{\infty}\cup B_0$ we expect these microstates
to reside on both $B_0$ and $B_\infty$.  However, it is well known that
the action in (\ref{gauge-fixed action}) is divergent due to the integral
over $B_\infty$, and that the {\it physical} (or regularized) action
is obtained by subtracting a suitable vacuum 
contribution \cite{Brow,Mann,Hawk}.
It turns out that
performing an exactly analogous subtraction of a (similarly divergent)
vacuum contribution to the total entropy leaves only the 
entropy contribution
from $B_0$.  Thus, the physically relevant microstates 
are present on $B_0$ only. 

To count these microstates is a formidable task.
At the classical level it can be shown that the reduced phase space
is the space of maps from the bifurcation surface
(more precisely, a constant Euclidean time slice, $\Sigma_\tau$,
of $B_0 =S^1 \times\Sigma$)
into the oriented Grassmann manifold 
$SO(d)/(SO(2)\times SO(d-2))$.  (The quotient by $SO(2)\times SO(d-2)$
reflects that the boundary theory considers as physical only the
orientation of an observer's frame {\it relative to the bifurcation
surface}.)  It is well known that such maps are intimately connected
with the theory of characteristic classes, in particular the Euler
number of $\Sigma_\tau$ 
(an intriguing example can be found in \cite{Cher}).
Furthermore, the scale of the reduced phase space is set by $A$,
the area of $\Sigma_\tau$ in Planck units.  So at the semi-classical
level, where the number of microstates is identified with the volume
of the reduced phase space, these two observations are suggestive
of the generalized entropy formula (\ref{generalized entropy}).
At the quantum level the numerical factor relating $A$ to $S$ turns out
(at least in the $d=3$ case) to be a function of the
parameter(s) labeling the unitary irreducible highest weight representations 
of the infinite dimensional Lie algebra
corresponding to the group of maps from $\Sigma_\tau$ into $SO(d)$
(subject to certain constraints).
(Roughly speaking, demanding the correct entropy formula fixes the
choice of representation.)
Interest in such representations originated in the study of 
anomalous gauge theories with chiral fermions, but more recently
has seen applications in $p$-branes and the study of non-perturbative
effects in string theory; unfortunately at present there exists only an
incomplete understanding of the mathematics 
involved (see, e.g., \cite{Mick,Ferr}).
In fact there are numerous other hints (not mentioned for lack of space)
that the boundary theory here has hidden connections with string theory,
and might eventually bridge the gap between an approach to microstates
based on standard general relativity and the string theory results
mentioned at the beginning.

In short, the ideas introduced here represent a 
concrete realization of a speculation by
Carlip and Teitelboim in which they suggest that the numerical factor
relating the entropy of a Euclidean black hole to the area of its 
bifurcation sphere might have its origin in microstates living on the
boundary of a thickened bifurcation sphere \cite{Carl2}.
Finally, we wish to emphasize that the counting of microstates here is
independent of the length scale associated with the
``thickening'' of the bifurcation surface: it is not ``physics at the
Planck scale'' on a horizon interpreted as a tangible 
boundary, but rather ``physics at no scale'' on a manifold ($B_0$) whose
{\it raison d'\^{e}tre} is topological in nature.

\clearpage

\vspace{1.5ex}
\begin{flushleft}
\large\bf Acknowledgments
\end{flushleft}
\noindent
R.J.E. would like to thank
Gabor Kunstatter, Jack Gegenberg, and Steven Carlip for stimulating
discussions and insightful criticisms.
This work was supported by grants from the Natural Sciences and
Engineering Research Council of Canada.

\end{document}